\documentclass[twocolumn,showpacs,preprintnumbers]{revtex4}
\usepackage{amsmath}
\usepackage{graphicx}
\usepackage{dcolumn}
\usepackage{bm}

\input{tcilatex}
\begin{document}

\title{Laboratory-scale superconducting mirrors for gravitational microwaves}
\author{Raymond Y. Chiao$^{\text{1,2}}$ and Stephen J. Minter$^{\text{2}}$}
\affiliation{$^{\text{1}}$University of California at Merced, School of Engineering and $%
^{\text{2}}$School of Natural Sciences, P.O. Box 2039, Merced, CA, 95344, USA}
\email{rchiao@ucmerced.edu; sminter2@ucmerced.edu}
\author{Kirk Wegter-McNelly}
\affiliation{Boston University, School of Theology, 745 Commonwealth Avenue, Boston, MA,
02215, USA}
\email{kwm@bu.edu}
\date{March 26a, 2009}

\begin{abstract}
When a gravitational wave at microwave frequencies impinges on a thin, type
I superconducting film, the radical delocalization of the film's negatively
charged Cooper pairs, which is due to the Uncertainty Principle, causes them
to undergo \textit{non-geodesic} motion relative to the \textit{geodesic}
motion of the decohered, positively charged ions in the film's lattice,
which is due to the Equivalence Principle. The ensuing charge separation
leads to a virtual plasma excitation. This \textquotedblleft
Heisenberg-Coulomb effect\textquotedblright\ enormously enhances the
interaction of a gravitational wave with a superconductor relative to that
of normal matter, so that the wave will be reflected even from a very thin
superconducting film. This result is presented using the BCS theory and a
superconducting plasma model.
\end{abstract}

\maketitle

Experiments at the frontiers of quantum mechanics and gravitation are rare 
\cite{Tajmar}\cite{Saulson}. In this Letter, we explore a prediction that
can eventually lead to experimental tests for the claim \cite%
{mirrors-PhysicaE} that two-dimensional superconducting films whose
thickness is less than the London penetration depth can specularly reflect
not only electromagnetic (EM) microwaves, as has been experimentally
demonstrated \cite{Glover-and-Tinkham}, but also gravitational (GR)
microwaves. We start with the fact that Einstein's field equations lead, in
the limits of \emph{weak }GR fields and \emph{nonrelativistic} matter, to
Maxwell-like equations \cite{Wald}, which in turn lead to boundary
conditions for gravitational fields at the surface of a superconducting film
that are homologous to those of electromagnetic fields. We find that the
response of a superconductor to a GR microwave is enhanced, relative to the
response of a normal conductor to the same wave, by the ratio of the
electrical force to the gravitational force between two electrons,%
\begin{equation}
\frac{e^{2}}{4\pi \varepsilon _{0}Gm_{e}^{2}}=4.2\times 10^{42}\text{ ,}
\label{Ratio of electrical to gravitational forces}
\end{equation}%
where $e$ is the electron charge, $m_{e}$ is the electron mass, $\varepsilon
_{0}$ is the permittivity of free space, and $G$ is Newton's constant. It is
the enormity of this pure number that leads to the reflection of a GR
microwave by a laboratory-scale superconductor.

Neglecting in a first approximation the weak interatomic forces that bind
normal matter together, the interaction between a GR microwave and a \emph{%
normal }metallic film, whose lateral dimensions are large compared to the
wavelength, can be modeled by treating the film's ions and normal electrons
as freely floating, non-interacting \textquotedblleft dust
particles\textquotedblright\ undergoing free-fall motion along
decoherence-induced \cite{Zurek} trajectories (i.e., geodesics). In this
approximation, the film \emph{cannot }interact energetically with a GR
microwave because each particle must, according to the Equivalence
Principle, remain at rest with respect to its local, co-moving, and
freely-falling inertial frame. Since there can be no energetic interaction
with the wave, mass currents cannot be generated locally within the film
without violating energy conservation.

This\ approximation fails in a superconductor since its Cooper pairs do not
possess decoherence-induced trajectories \cite{mirrors-PhysicaE}. According
to the Bardeen--Cooper--Schrieffer (BCS) theory of superconductivity, each
Cooper pair is in a zero-momentum eigenstate, relative to the center of mass
of the system, when the system is in the BCS ground state \cite%
{Glover-and-Tinkham}. This implies that their positions inside the
superconductor are completely \emph{uncertain}, i.e., that their
trajectories are completely \emph{delocalized}. Thus Cooper pairs cannot
undergo free fall along with the ions and normal electrons, so that an
application of the Equivalence Principle to the motion of the pairs is
precluded by the Uncertainty Principle. Quantum delocalization causes the
Cooper pairs of a superconductor to undergo \emph{non-geodesic} motion
relative to the \emph{geodesic }motion of its ionic lattice, which leads to
quantum supercurrents that contain kinetic energy extracted from the wave.

The generation of supercurrents within a superconductor by a GR microwave
has an important consequence, namely, the electrical polarization of the
superconductor. The Coulomb force of attraction, resulting from oppositely
signed charges accumulating along the edges of the superconductor, leads to
an effective Hooke's law restoring force that strongly opposes the
gravitational force of the incoming wave. The enormous back-action of the
Coulomb force on the motion of the Cooper pairs, which we refer to as the
\textquotedblleft Heisenberg-Coulomb effect,\textquotedblright\ greatly
enhances the mass supercurrents generated by the wave, so that they become
strong enough to produce reflection.

In the EM sector, the interaction of radiation with a film having large
lateral dimensions compared to the wavelength and a sufficiently small
thickness $d$ can be modeled using \textquotedblleft
lumped-circuit\textquotedblright\ concepts such as resistance, inductance,
etc., of an infinitesimal square element of the film. In the case of a film
with an arbitrary, frequency-dependent complex conductivity $\sigma (\omega
)=\sigma _{1}(\omega )+i\sigma _{2}(\omega ),$ the reflectivity ${\mathcal{R}%
}$ is given by \cite{mirrors-PhysicaE}%
\begin{equation}
{\mathcal{R}}=\left\{ \left( 1+\frac{\sigma _{1}}{\sigma _{1}^{2}+\sigma
_{2}^{2}}\frac{2}{Z_{0}d}\right) ^{2}+\left( \frac{\sigma _{2}}{\sigma
_{1}^{2}+\sigma _{2}^{2}}\frac{2}{Z_{0}d}\right) ^{2}\right\} ^{-1}\text{ ,}
\label{EM-reflectivity}
\end{equation}%
where $Z_{0}$ is the EM characteristic impedance of free space.

Consider now a superconducting film whose thickness $d$ is much smaller than
the coherence length $\xi _{0}$ and the London penetration depth $\lambda _{%
\text{L}}$ of the material. As Tinkham has noted \cite{Glover-and-Tinkham},
the dissipative part of the conductivity of such a film $\sigma _{1\text{s}}$
goes \emph{exponentially} to zero as $T\rightarrow 0$ in response to a
driving wave whose quanta carry energies less than $\hbar \omega _{\text{gap}%
}=2\Delta (0)\cong 3.5k_{\text{B}}T_{\text{c}}$, where $\Delta (0)$
(henceforward abbreviated as $\Delta )$ is the gap energy per electron of
the BCS theory at $T=0$. This exponential suppression is due to the
\textquotedblleft freezing out\textquotedblright\ of the film's normal
electrons through the Boltzmann factor $\exp \left( -\Delta /k_{\text{B}%
}T\right) $ as $T\rightarrow 0$. On the other hand, the film's
non-dissipative conductivity $\sigma _{2\text{s}}$ rises asymptotically to
some finite value in the same limit \cite{Glover-and-Tinkham}. The behavior
of $\sigma _{2\text{s}}$ is due to the film's inductive reactance $X_{\text{L%
}}$, which in turn arises from its inductance (per square element of the
film) $L$, according to the relations $X_{\text{L}}=1/(\sigma _{2\text{s}%
}d)=\omega L$.

For a superconducting film at temperatures sufficiently near $T=0$ and for
frequencies lower than $\omega _{\text{gap}}$, the ohmic dissipation of the
film will be exponentially suppressed by the Boltzmann factor, so that one
can, to a good approximation, set $\sigma _{1\text{s}}=0$ and rewrite (\ref%
{EM-reflectivity}), as well as identify the \textquotedblleft
roll-off\textquotedblright\ frequency $\omega _{\text{r}}$ (at which the
reflectivity ${\mathcal{R}}_{\text{s}}$ drops to $50\%$), as%
\begin{equation}
{\mathcal{R}}_{\text{s}}=\left\{ 1+\left( 2\frac{X_{\text{L}}}{Z_{0}}\right)
^{2}\right\} ^{-1}\text{ \ and \ }\omega _{\text{r}}=\frac{Z_{0}}{2L}\text{ .%
}  \label{EM-reflectivity-imped-ratio}
\end{equation}

The inductance $L$ of the film has two parts: a magnetic part $L_{\text{m}}$
due to the magnetic fields created by the electrical supercurrents carried
by the Cooper pairs, and a kinetic part $L_{\text{k}}$ due to the Cooper
pairs' inertial mass, by which they oppose the accelerating force of the
external electric field \cite{Glover-and-Tinkham}. We have shown that $L_{%
\text{m}}$ is negligible compared to $L_{\text{k}}$ for a superconducting
film, so that $L\approx L_{\text{k}}$ \cite[Appendix A]{mirrors-PhysicaE}.

The BCS relation between the imaginary part of a superconducting film's
complex conductivity $\sigma _{\text{2s}}$ (when $T$ $\ll T_{\text{c}}$ and $%
\omega \ll \omega _{\text{gap}}$) and the film's normal conductivity $\sigma
_{\text{n}}$, as well as the Drude expression for the film's normal
conductivity $\sigma _{\text{n}}$ \cite[footnote 14]{mirrors-PhysicaE}, then
lead to the following expression for the film's inductance:%
\begin{equation}
L_{\text{k}}=\frac{1}{\omega \sigma _{2\text{s}}d}=\frac{1}{d^{2}}\cdot 
\frac{\hbar v_{\text{F}}}{\pi \Delta }\cdot \frac{m_{\text{e}}}{n_{\text{e}%
}e^{2}}\text{ ,}  \label{L_k}
\end{equation}%
where $v_{\text{F}}$ is the Fermi velocity and $n_{\text{e}}$ is the number
density of electrons. The $1/d^{2}$ factor in (\ref{L_k}) indicates a
inverse-square dependence on the film's thickness, whereas the presence of $%
\hbar v_{\text{F}}/\pi \Delta \equiv \xi _{0}$ implies a linear dependence
on the coherence length. The $m_{\text{e}}/(n_{\text{e}}e^{2})$ term is
related to the plasma frequency $\omega _{\text{p}}$.

In the limit of $\omega \ll \omega _{\text{p}},$ the plasma skin depth $%
\delta _{\text{p}}$ is given by%
\begin{equation}
\delta _{\text{p}}=\frac{c}{\omega _{\text{p}}}=\sqrt{\frac{m_{\text{e}}}{%
\mu _{0}n_{\text{e}}e^{2}}}\text{ ,}  \label{EM-skin-depth}
\end{equation}%
which is identical to the London penetration depth $\lambda _{\text{L}}$.
Since $L_{\text{k}}=\mu _{0}l_{\text{k}}$, where $l_{\text{k}}$ is the
characteristic length scale associated with the film's kinetic inductance,
it follows that%
\begin{equation}
l_{\text{k}}=\xi _{0}\left( \delta _{\text{p}}/d\right) ^{2}\text{ .}
\label{l_k}
\end{equation}%
The roll-off frequency $\omega _{\text{r}}$ then becomes%
\begin{equation}
\omega _{\text{r}}=Z_{0}/(2L)=c/(2l_{\text{k}})\text{ .}  \label{omega_k}
\end{equation}%
In the case of an EM wave with frequency $\omega $ and a superconducting
film at temperatures sufficiently near $T=0,$ the possibility of specular
reflection depends only on the ratio of the speed of light $c$ to the
kinetic inductance length scale $l_{\text{k}}$. For a superconducting film
made of lead used in \cite{Glover-and-Tinkham}, $l_{\text{k}}\approx 30\mu $%
m and $\omega _{\text{r}}\approx 2\pi \times (840$ GHz$)$.

Let us now turn to the case of incident GR waves on these films. Wald \cite[%
Section 4.4]{Wald} has introduced a useful approximation scheme that leads
to a Maxwell-like representation of the Einstein equations of general
relativity. The resulting equations describe the coupling of weak GR fields
to nonrelativistic matter and lead to a wave equation for GR waves analogous
to the standard wave equation for EM waves. In the asymptotically flat
spacetime coordinate system of a distant inertial observer and in SI units,
the four Maxwell-like equations are 
\begin{subequations}
\label{Maxwell-like-eqs}
\begin{gather}
\mathbf{\nabla \cdot E}_{\text{G}}=-\frac{\rho _{\text{G}}}{\varepsilon _{%
\text{G}}}  \label{Maxwell-like-eq-1} \\
\mathbf{\nabla \times E}_{\text{G}}=-\frac{\partial \mathbf{B}_{\text{G}}}{%
\partial t}  \label{Maxwell-like-eq-2} \\
\mathbf{\nabla \cdot B}_{\text{G}}=0  \label{Maxwell-like-eq-3} \\
\mathbf{\nabla \times B}_{\text{G}}=\mu _{\text{G}}\left( -\mathbf{j}_{\text{%
G}}+\varepsilon _{\text{G}}\frac{\partial \mathbf{E}_{\text{G}}}{\partial t}%
\right)\text{ ,}  \label{Maxwell-like-eq-4}
\end{gather}%
where the gravitational analog of the electric permittivity of free space is 
$\varepsilon _{\text{G}}=1/(4\pi G)=1.2\times 10^{9}$ SI units and the
gravitational analog of the magnetic permeability of free space is $\mu _{%
\text{G}}=4\pi G/c^{2}=9.3\times 10^{-27}$ SI units.

The field $\mathbf{E}_{\text{G}}$ in these equations is the \emph{gravito}%
-electric field, which is to be identified with the local acceleration $%
\mathbf{g}$ of a test particle produced by the mass density $\rho _{\text{G}}
$, in the Newtonian limit of general relativity. The field $\mathbf{B}_{%
\text{G}}$ is the \emph{gravito}-magnetic field produced by the mass current
density $\mathbf{j}_{\text{G}}$ and by the gravitational analog of the
Maxwell displacement current density $\varepsilon _{\text{G}}\partial 
\mathbf{E}_{\text{G}}/\partial t$ of the Ampere-like law (\ref%
{Maxwell-like-eq-4}). This magnetic-like field is a generalization of the
Lense-Thirring field of general relativity.

An important property that follows from (\ref{Maxwell-like-eqs}) is the 
\textit{gravitational} characteristic impedance of free space \cite%
{mirrors-PhysicaE}: 
\end{subequations}
\begin{equation}
Z_{\text{G}}=\sqrt{{\mu _{\text{G}}}/{\varepsilon _{\text{G}}}}=\mu _{\text{G%
}}c=2.8\times 10^{-18}\text{ SI units.}  \label{Z_G}
\end{equation}%
This quantity is a characteristic of the vacuum and is independent of any of
the properties of matter \textit{per se}, i.e., it is a property of
spacetime itself. As with $Z_{0}$ in the EM sector, $Z_{\text{G}}$ will play
a central role in all GR radiation coupling problems. The impedance of any
material object needs to be much smaller than this extremely small quantity
before the object can reflect any significant portion of an incident GR
wave. In other words, conditions must be highly unfavorable for dissipation
into heat.

Since a given Cooper pair carries both mass and charge, a tidal, quadrupolar
pattern of mass and electrical supercurrents will be induced inside a
superconducting film by an incident GR microwave, from the perspective of a
local, freely falling observer who is near the surface of the film and who
is located anywhere other than at the film's center of mass. For weak GR
wave amplitudes, the mass supercurrents will be describable by the linear
relationship 
\begin{equation}
\mathbf{j}_{\text{G}}(\omega )\mathbf{=}\sigma _{\text{s,G}}(\omega )\mathbf{%
E}_{\text{G-inside}}(\omega )\text{ ,}  \label{Gravitational-Ohm's-law}
\end{equation}%
where $\mathbf{j}_{\text{G}}(\omega )$ is the mass supercurrent density at
frequency $\omega $, $\sigma _{\text{s,G}}(\omega )=\sigma _{\text{1s,G}%
}(\omega )+i\sigma _{\text{2s,G}}(\omega )$ is the complex mass-current
conductivity of the film at the frequency $\omega $ in its linear response
to the fields of the incident GR wave, and $\mathbf{E}_{\text{G-inside}%
}(\omega )$ is the driving gravito-electric field inside the film at
frequency $\omega $.

On the basis of the similarity of the Maxwell to the Maxwell-like equations,
the identicality of the boundary conditions that follow from these
equations, and the linearity of (\ref{Gravitational-Ohm's-law}), we are led
to the following two expressions in the GR sector for the reflectivity and
the roll-off frequency, which are analogous to (\ref%
{EM-reflectivity-imped-ratio}) in the EM sector:%
\begin{equation}
{\mathcal{R}}_{\text{G}}=\left\{ 1+\left( 2\frac{X_{\text{L,G}}}{Z_{\text{G}}%
}\right) ^{2}\right\} ^{-1}\text{ \ and \ }\omega _{\text{r,G}}=\frac{Z_{%
\text{G}}}{2L_{\text{G}}}\text{ .}  \label{GR-refl-cond}
\end{equation}%
As in the EM case, we can neglect the contribution of the gravito-magnetic
inductance $L_{\text{m,G}}$ to the overall gravitational inductance $L_{%
\text{G}}$ \cite[Appendix A]{mirrors-PhysicaE}, so that $L_{\text{G}}\approx
L_{\text{k,G}}=\mu _{\text{G}}l_{\text{k,G}}$. Then the roll-off frequency $%
\omega _{\text{r,G}}=c/(2l_{\text{k,G}})$ depends, as before, on the ratio
of the speed of light $c$ to a single length scale -- in this case, the 
\emph{gravitational }kinetic inductance length scale $l_{\text{k,G}}$.

For the moment, let us assume that the coupling of Cooper pairs to a GR wave
depends \emph{solely} on the gravitational mass $2m_{\text{e}}$ of the
Cooper pairs, i.e., that their electrical charge $2e$ is irrelevant to the
gravitational plasma skin depth and thus to the gravitational kinetic
inductance length scale. This is tantamount to treating the Cooper pairs as
if they were neutral particles. We can then obtain the \textquotedblleft
gravitational\textquotedblright\ version of the kinetic inductance length
scale $l_{\text{k,G}}$ by making the replacement $e^{2}/4\pi \varepsilon
_{0}\rightarrow Gm_{\text{e}}^{2}$ in (\ref{EM-skin-depth}) and (\ref{l_k}),
which leads to%
\begin{equation}
l_{\text{k,G }}=\xi _{0}\left( \delta _{\text{p,G }}/d\right) ^{2}\text{ ,}
\label{l_kG-BCS}
\end{equation}%
where the \textquotedblleft gravitational\textquotedblright\ plasma skin
depth $\delta _{\text{p,G}}$ is given by%
\begin{equation}
\delta _{\text{p,G }}=\sqrt{\frac{1}{\mu _{\text{G}}n_{\text{e}}m_{\text{e}}}%
}\text{ }.  \label{plas-skin-dep-GR}
\end{equation}%
This expression for $\delta _{\text{p,G}}$ leads to an astronomically large
kinetic inductance length scale $l_{\text{k,G }}$ on the order of $10^{36}$
m for the superconducting lead films used in \cite{Glover-and-Tinkham},
which would appear to preclude any possibility of laboratory-scale
GR-microwave reflection. But this analysis is flawed, for it ignores the
\textquotedblleft Heisenberg-Coulomb effect,\textquotedblright\ i.e., the
enormous Coulomb forces within the film due to the electrical charge of its
delocalized Cooper pairs.

A quantum treatment of the motion of the Cooper pairs begins with the
probability current density $\mathbf{j}$ in non-relativistic quantum
mechanics, which is given by%
\begin{equation}
\mathbf{j}=\frac{\hbar }{2mi}(\psi ^{\ast }\nabla \psi -\psi \nabla \psi
^{\ast })\text{ },  \label{prob. curr. density j}
\end{equation}%
where $m$ is the mass of the nonrelativistic particle whose current is being
calculated (here $m=2m_{e}$) and $\psi $ is the wavefunction of the system
(here the Cooper pair's \textquotedblleft condensate
wavefunction\textquotedblright ). Following DeWitt \cite{DeWitt}, we use the
minimal coupling rule for the momentum operator%
\begin{equation}
\mathbf{p}\rightarrow \mathbf{p}-q\mathbf{A}-m\mathbf{h}\text{ ,}
\label{DeWitt's minimal coupling rule}
\end{equation}%
where $q=2e$, $m=2m_{e}$, $\mathbf{A}$ is the electromagnetic vector
potential, and $\mathbf{h}$ is the gravitational vector potential \cite%
{DeWitt}.

In the case of a superconducting film before the arrival of a GR microwave,
its Cooper pairs will be in a zero-momentum eigenstate where $\psi $ is
constant. According to the quantum adiabatic theorem and first-order
perturbation theory, this wavefunction must remain unchanged to lowest order
by any radiative perturbations arising from either $\mathbf{A}$ or $\mathbf{h%
}$ after the arrival of a wave whose frequency is less than the BCS gap
frequency. Defining the quantum velocity field induced by a GR wave as $%
\mathbf{v}=\mathbf{j/}\left( \psi ^{\ast }\psi \right) $, it then follows
from (\ref{prob. curr. density j}) and (\ref{DeWitt's minimal coupling rule}%
) that%
\begin{equation}
\mathbf{v}=-\frac{q}{m}\mathbf{A-h}\text{ }.  \label{velocity field}
\end{equation}%
By the radiation-gauge choice where $\mathbf{{E}=-\partial {A}/\partial }t$
and $\mathbf{E}_{\text{G}}\mathbf{=-\partial {h}/\partial }t$, the mass
current density source term in the Ampere-like law of the Maxwell-like
equations (\ref{Maxwell-like-eq-4}) becomes, upon taking a time derivative
of the above equation and assuming a common $\text{exp}(-i\omega t)$ time
dependence for all fields,%
\begin{equation}
\mathbf{j}_{\text{G}}=nm\mathbf{v}=i\frac{n}{\omega }\left( q\mathbf{E+}m%
\mathbf{E}_{\text{G}}\right) \text{ ,}  \label{j_G vs. E and E_G}
\end{equation}%
where $n$ is the number density of Cooper pairs. This result implies that
the total force acting on a given Cooper pair is $\mathbf{F}_{\text{tot}}=q%
\mathbf{E}+m\mathbf{E}_{\text{G}}$.

When a superconductor is operating in its \emph{linear response regime }in
the presence of a weak incident GR microwave, the direct proportionalities $%
\mathbf{F}_{\text{tot}}\propto \mathbf{E\propto \mathbf{E}_{\text{G}}}$ will
hold. We can therefore define a proportionality constant $\Xi $, such that $%
\mathbf{F}_{\text{tot}}=\Xi q\mathbf{E}$. If we then derive a \emph{modified}
plasma frequency $\omega _{\text{p}}^{\prime }$, allowing for the
possibility of an extremely small correction arising from the gravitational
attraction between electrons, we find that%
\begin{equation}
\omega _{\text{p}}^{\prime 2}=\Xi \frac{nq^{2}}{m\varepsilon _{0}}\text{ .}
\label{mod plas freq in terms of Ksi}
\end{equation}

In light of the proportionalities introduced above, the relationship between
the $\mathbf{E} $ and $\mathbf{E}_{\text{G}}$ fields inside a
superconducting film when $\Xi \neq 1,$ i.e., when the extremely weak
gravitational forces within the film are taken into account, is given by%
\begin{equation}
\mathbf{E}=\frac{1}{\Xi -1}\frac{m}{q}\mathbf{E}_{\text{G}}\text{ }.
\label{E_in_terms_of_E_G}
\end{equation}%
Substituting this result into (\ref{j_G vs. E and E_G}), one finds that the
mass conductivity of the film $\sigma _{\text{G}}$ is given by%
\begin{equation}
\sigma _{\text{G}}=i\left( \frac{\Xi }{\Xi -1}\right) \frac{nm}{\omega }%
\text{ .}  \label{sigma_G in terms of Ksi}
\end{equation}%
Note that $\sigma _{\text{G}}$ can become extremely large when $\Xi
\rightarrow 1$, and therefore that $\mathbf{j}_{\text{G}}$\ can also become
extremely large.

To determine $\Xi $, we begin by multiplying (\ref{velocity field}) by $nq$
and taking a time derivative:%
\begin{equation}
\frac{\partial }{\partial t}\mathbf{j}_{\text{e}}=\frac{\partial (nq\mathbf{v%
})}{\partial t}=\frac{nq^{2}\mathbf{E}}{m}+nq\mathbf{E}_{\text{G}}\ ,
\label{Current EOM}
\end{equation}%
which can be evaluated at a point $P$ along the edge of the superconductor
where the ionic lattice abruptly ends and the vacuum begins. We assume that
the incident radiation field that excites the superconducting plasma is
tightly focused onto a diffraction-limited Gaussian-beam spot size located
at the center of the plasma, whose lateral dimensions are much larger than a
wavelength of the incident radiation. We also assume that the radiative
excitation is impulsive in nature, so that the plasma can oscillate freely
after the radiation is suddenly turned off. Taking the divergence of (\ref%
{Current EOM}), one obtains the simple harmonic equation of motion%
\begin{equation}
\frac{\partial ^{2}}{\partial t^{2}}\rho _{\text{e}}+\frac{nq^{2}}{%
m\varepsilon _{0}}\rho _{\text{e}}-\frac{nm}{\varepsilon _{\text{G}}}\rho _{%
\text{e}}=\frac{\partial ^{2}}{\partial t^{2}}\rho _{\text{e}}+\omega _{%
\text{p}}^{\prime 2}\rho _{\text{e}}=0\text{ ,}
\end{equation}%
where the modified plasma frequency is given by%
\begin{equation}
\omega _{\text{p}}^{\prime 2}=\frac{nq^{2}}{m\varepsilon _{0}}\left( 1-\frac{%
m^{2}}{q^{2}}\frac{Z_{\text{G}}}{Z_{0}}\right) \text{ .}
\label{modi_plas_freq_Z_0_Z_G}
\end{equation}

Comparing (\ref{modi_plas_freq_Z_0_Z_G}) with (\ref{mod plas freq in terms
of Ksi}), we see that%
\begin{equation}
\Xi =1-\frac{m^{2}}{q^{2}}\frac{Z_{\text{G}}}{Z_{0}}=1-\frac{4\pi
\varepsilon _{0}Gm_{\text{e}}^{2}}{e^{2}}\text{ ,}  \label{Ksi}
\end{equation}%
since $Z_{G}=4\pi G/c$ and $Z_{0}=1/(\varepsilon _{0}c)$. Thus, the
parameter $\Xi $ differs from unity by an extremely small amount equal to
the reciprocal of the ratio of the forces for electrons given by (\ref{Ratio
of electrical to gravitational forces}). From this it follows that the \emph{%
corrected }gravitational kinetic inductance length scale $l_{\text{k,G}%
}^{\prime }$ is given by the microscopic quantity%
\begin{equation}
l_{\text{k,G}}^{\prime }=d\left( \delta _{\text{p}}/d\right) ^{2}\text{ }
\label{l_kG-plasma-corr-factor}
\end{equation}%
instead of the astronomically large length scales given by (\ref{l_kG-BCS})
and (\ref{plas-skin-dep-GR}).

Note that (\ref{l_kG-plasma-corr-factor}) is just the EM kinetic inductance
length scale $l_{\text{k}}$ given by (\ref{l_k}) apart from a factor on the
order of unity, i.e., $d/\xi _{0}$, which arises from the absence of the BCS
coherence length scale $\xi _{0}$ in the plasma model. But this result
already shows that the Heisenberg-Coulomb effect reduces the GR kinetic
inductance length scale $l_{\text{k,G}}$ by $42$ orders of magnitude,
originating from (\ref{Ratio of electrical to gravitational forces}), to the
level of the EM kinetic inductance length scale $l_{\text{k}}$, and thereby
increases the magnitude of the GR roll-off frequency $\omega _{\text{r,G}}$
by the same factor, to the level of the EM roll-off frequency $\omega _{%
\text{r}}$. We therefore conclude that laboratory-scale superconducting
mirrors for GR microwaves exist.

\end{document}